\documentclass[twocolumn,aps,prb,superscriptaddress,longbibliography]{revtex4-2}
\usepackage{graphicx}
\usepackage{color}
\usepackage{amsmath}
\usepackage[export]{adjustbox}
\usepackage{enumitem}
\usepackage{amssymb}
\usepackage{hyperref}
\usepackage[normalem]{ulem}
\usepackage{adjustbox}
\usepackage{float}

\hypersetup{
	colorlinks = true,
	linkcolor = blue,
	citecolor = blue,
	urlcolor  = blue,
}

\newcommand{\be}{\begin{equation}}
	\newcommand{\ee}{\end{equation}}

\newcommand{\bea}{\begin{eqnarray}}
	\newcommand{\eea}{\end{eqnarray}}

\renewcommand{\vec}[1]{{\boldsymbol #1}}
\renewcommand{\epsilon}{\varepsilon}

\def\nn{\nonumber\\}

\begin{document}


\title{Size and Impurity Effects on Scattering of Valley Hall Modes in Gate-Defined Bilayer Graphene Superlattices}
 
	\date{\today}

    \author{Marcus N. Kanestrøm}
    \affiliation{Kavli Institute of Nanoscience, Delft University of Technology, 2628 CJ Delft, the Netherlands}

 \author{Antonio L. R. Manesco}
 \affiliation{Center for Quantum Devices, Niels Bohr Institute,
University of Copenhagen, DK-2100 Copenhagen, Denmark}
    
	\author{D. O. Oriekhov}
    \email{d.oriekhov@tudelft.nl}
	\affiliation{Kavli Institute of Nanoscience, Delft University of Technology, 2628 CJ Delft, the Netherlands}
	
	\begin{abstract}
    In the present paper we perform a tight-binding simulation of gate-defined islands in Bernal bilayer graphene (BLG). The inversion of the gap sign on the boundaries of the islands creates topologically-protected valley Hall modes. We focus on the specific questions of whether the valley Hall modes around such islands could serve as a host for quantum walks or simulate weakly coupled systems, and how their tunneling is affected by in-gap impurities. In addition, we discuss the effect of misalignment of top and bottom gate patterns on the tunneling properties between islands. Our main results show that resonant tunneling via an impurity enhances overlap between superlattice islands, while misalignment does not break topological protection over a wide parameter regime. In addition, we study the two-island geometry and show that it is possible to leverage suppressed scattering to place islands more densely on a single sample.   
	\end{abstract}
	\maketitle

\section{Introduction}


Topologically protected edge modes in quantum Hall bar devices and in topological insulators are attracting a great interest for their possible quantum information applications. Their localized structure, symmetry-protected uni-directional movement and the high robustness against disorder make them natural candidates for the so-called flying qubits \cite{divincenzo1996topicsquantumcomputers,Bordone2019SemiconductorScienceQuantumComputingQuantum}. A number of theoretical works predicted that bilayer graphene with a sign-alternating gap is one of the simplest lattice structures that could host such modes.
Very recently experimental findings confirmed this expectation. In this work, we study the scattering properties of these modes between several separate channels and on charged impurities, and describe the possible quantum information applications.

Bernal bilayer graphene consists of two sheets of graphene with the top sheet having one sublattice directly above the atoms of lower sheet, and another sublattice being above the centers of hexagons in lower sheet \cite{McCannKoshino2013}. Due to its bilayer structure, the band gap could be tuned over a wide range by applying a voltage difference between layers \cite{Zhang2009}. Gapped bilayer graphene is not a Chern insulator itself, as it maintains time-reversal symmetry \cite{McCannKoshino2013}. But the states near the $K$ and $K'$ valleys in momentum space individually break time-reversal symmetry. Theoretical works have shown, that in bilayer graphene this leads to a non-zero valley Chern number \cite{Martin2008TopologicalConfinement,Zhang2013PNAS,Alden2013PNAS,Vaezi2013PRX}. This topological invariant is an analogue of a Chern number, but its integration is restricted to a part of momentum space. 
The value of valley Chern number equals to $\pm 1$ for the upper and lower band in K valley, and inverse sign in $K'$ valley. This value guarantees the existence of one edge mode on the terminated side, or two edge modes in the region with switched gap sign according to bulk-boundary correspondence \cite{Martin2008TopologicalConfinement,Zhang2013PNAS,Alden2013PNAS,SanJose2013Helic,Tsim2020twist}. 
The difficulty in observing these edge modes at terminated boundary of graphene comes from the existence of a rich spectrum of non-topological modes due to ribbon lattice imperfections.
Thus, to observe these modes, one needs to engineer a device with alternating gap structure inside the bulk of large sample.

Historically, the first experimental evidence of the possible existence of such modes appeared in relaxed minimally-twisted bilayer graphene \cite{Rickhaus2018,Huang2018PRL_top, Xu2019NatComm,Mahapatra2022}. Specifically, the STM imaging techniques \cite{Rickhaus2018,Xu2019NatComm} confirmed the localization of the found modes within domain walls. The experiments in Ref.\cite{Verbakel2021} further provided the evidence of the good level of valley coherence in such domain walls. But, the later progress in theoretical understanding of physics of domain walls has shown that the deformation picture makes the real spectrum way more complicated \cite{Ceferino2023}. The main origin is the appearance of strong pseudomagnetic field, that creates extra non-chiral modes confined to domain wall. The experimental setups with deformation domain walls still do not allow the clear identification and separation of chiral and non-chiral modes in twisted bilayer graphene samples.


At the same time, the possible existence of chiral states in domain walls, connected into coherent networks, has a number of theoretical implications. The one is related to the above-mentioned flying qubits \cite{divincenzo1996topicsquantumcomputers}, where already one channel could serve as connector in a quantum computer. Another set of  applications is based on the structure of networks. A number of works \cite{Efimkin2018PRB,DeBeule2020,DeBeule2021PRB_floquet,DeBeule2021PRB_network,Wittig2023PRB} described these networks phenomenologically and proposed a range of phenomena to be observed, such as Aharonov-Bohm oscillations, valley-locked currents and valley splitters, and Floquet-type spectrum.
The Ref.\cite{Vakhtel2022PRBL} proposed an idea 
to use such networks as quantum walks - a specific 
quantum algorithm implemented on a crystal - to 
observe Bloch oscillations. The main caveat behind 
a number of these theoretical works in the hardness 
of obtaining phenomenological parameters from 
the ab-initio microscopic calculations. The origin 
of this problem stays in the enormous 
computational complexity of simulating relaxed multi-domain Moire lattice.

The problems of full-scale lattice simulations and the experimental separation of chiral modes in domain walls are naturally solvable in regular Bernal bilayer graphene. The only experimental requirement is an outstanding quality of both sample and gates, which was reportedly reached in Ref.\cite{Huang2024HighTemperatureQVH}. This motivates to study the scattering of such modes theoretically in various configurations, to obtain the missing link between theoretical network models and the microscopic parameters.  In particular, the corresponding studies were already done for the properties of valley Hall modes around single gate-defined island in Ref.\cite{Xavier2010} analytically and in Refs.\cite{Benchtaber2021,Benchtaber2022pssb,Jaskolski2026} numerically for islands of small sizes and certain shapes. The scattering specifically between the states in two or more gate-defined domain walls were studied in Ref.\cite{Benchtaber2021scatt} for two parallel domain walls and in Ref.\cite{Luna2025SciPost} for the valley splitter structures. In the latter, the Y-shaped connection between standard conducting channels and gap-sign-switching domain walls was predicted to be capable of separating valleys with high probability.

\begin{figure}
    \centering
    \includegraphics[scale=1.0]{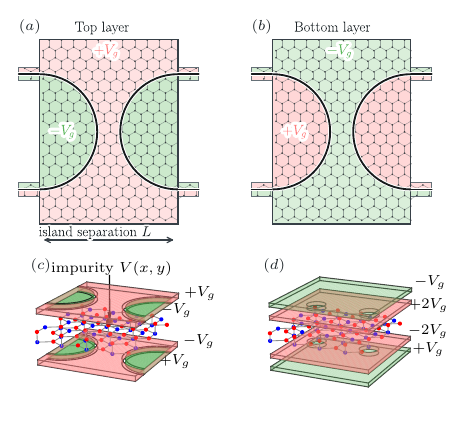}
    \caption{(a),(b) The impurity free BLG device and the associated layer potentials. Attached leads are the extended portions of the rectangular shape on both the left and right side, where the domain wall enters. (a) The top graphene layer, where the half circle cutouts have a negative bias and the bulk scattering region has a positive bias. (b) The bottom graphene layer, which is a mirror copy of the top layer, but this time the interlayer bias is opposite to that of the top layer. 
    Panels (c) and (d) represent a 3D view of devices studied in this paper. The (c) setup is used to study the scattering properties of helical modes with and without impurity potential $V(x,y)$. The two island setup in (d) is used to compare the S-matrix tunneling amplitudes with the scale of level splitting.
     In addition, panel (d) suggests a way to realize multi-island device via aligned holes in middle gates.}
    \label{fig:top_down_device_schematic}
\end{figure}

In the present work, we focus on the regime of a fully gapped system, where the helical modes could interact only via in-gap tunneling.  The schematic device picture with potential patterns of top and bottom gates is shown in Fig.\ref{fig:top_down_device_schematic}. We study the cases of strong and weak tunneling between two closely-placed channels, and identify the respective parameter dependencies. The first regime could be potentially used for quantum walk or other quantum algorithm applications. We treat this as a system of closely-placed gate-defined helical quantum dots. For the second regime we discuss the possible molecular simulator application. Important to note, that both regimes could be also viewed as a limiting factors on coherence of single mode. In other words, if the mode should maintain coherence and no loss of its quantum state, the above-mentioned scattering sets hard limits on the possible density of quantum dots on one sample.
In addition, we study the regime of scattering on charged impurity potential. We show that the resonant scattering on a bound state preserves valley index up to high prevision. All numerical calculations are done with Kwant \cite{groth2014kwant}  package.

The paper is organized as follows: in Sec.\ref{sec:helical-modes} we describe the gate-defined domain walls and the main properties of helical modes. In addition, we discuss the scaling of decay length of modes with the gap parameter. In the next sections these properties of modes are used to optimize  complexity of numerical simulation.
In Sec.\ref{sec:scattering-between-islands} we introduce the system structure and the central scattering problem. The main results for scattering matrices between helical states in different configurations are presented. In Sec.\ref{sec:misalignment} we further expand these results to the more realistic geometry of misaligned top and bottom gate patterns. Finally, in Sec.\ref{sec:quantum-dot-simulator} we provide an  example of two chiral islands and the level splitting when they are at close distance. At certain energy levels the splitting is almost absent, which corresponds to protection of chiral quantum dot islands from interacting with each other via tunneling.

\section{Helical domain wall states in Bernal bilayer graphene}
\label{sec:helical-modes}

\subsection{Tight-binding model of Bernal bilayer graphene}

In the Bernal (AB) stacking configuration, atom $B_1$ lies directly above atom $A_2$, forming the dimer sites that are coupled by the strongest interlayer hopping parameter $\gamma_1$, while atoms $A_1$ and $B_2$ are the non-dimer sites. This stacking geometry plays a central role in determining the low-energy electronic properties of bilayer graphene. In particular, the application of an interlayer potential difference breaks inversion symmetry and opens a bulk band gap. When the sign of this interlayer potential changes across a domain wall, the valley Chern number changes accordingly, giving rise to topologically protected valley Hall chiral modes that are localized along the domain wall and propagate in opposite directions in the two inequivalent valleys.

The Bernal or AB stacking of two graphene layers is a stacking where the upper layer is rotated compared to the bottom one by $\pi/3$. The unit cell consists of four atoms that are labeled as $A_1$, $B_1$ for lower layer, and $A_2$, $B_2$ for upper layer. Their in-plane coordinates are given by $\boldsymbol{\tau}_{A_1}=(0,0)$, $\boldsymbol{\tau}_{B_1}=\frac{1}{3}(\mathbf{a}_1+\mathbf{a}_2)$, $\boldsymbol{\tau}_{A_2}=\frac{1}{3}(\mathbf{a}_1+\mathbf{a}_2)$, and $\boldsymbol{\tau}_{B_2}=(0,0)$. The primitive lattice vectors are defined as $\mathbf{a}_1=a\left(\frac{\sqrt{3}}{2},\frac{3}{2}\right)$ and $\mathbf{a}_2=a\left(-\frac{\sqrt{3}}{2},\frac{3}{2}\right)$, where $a=1.42\,\text{\AA}$ denotes the carbon-carbon bond length. Equivalently, the graphene lattice constant is $a_0=\sqrt{3}a=2.46\,\text{\AA}$. The separation between the two graphene layers is typically taken to be $d=3.35\,\text{\AA}$.  In the numerical simulations the z-coordinates of atoms are not used as they are constant and do not influence the in-plane scattering in the absence of ripples in the model.

The nearest neighbor tight-binding model of BLG in the second quantization is described by the following Hamiltonian:

\begin{align}
\hat{H} =  & \hat{H}_{0}+\hat{H}_{t}+\hat{H}_{\gamma_1}+\hat{H}_{\gamma_3}+\hat{H}_{\gamma_4}\nn
\hat{H}_0 = &
\sum_{l=1}^{2}\sum_{i}\epsilon^{A}_{l,i}\,\hat{a}^{\dagger}_{l,i}\hat{a}_{l,i}
+
\sum_{l=1}^{2}\sum_{i}\epsilon^{B}_{l,i}\,\hat{b}^{\dagger}_{l,i}\hat{b}_{l,i} \nn
\hat{H}_{t} = &- t \sum_{l=1}^{2}\sum_{\langle i,j\rangle}\left(\hat{a}^{\dagger}_{l,i} \hat{b}_{l,j} + \mathrm{h.c.}\right) \nn
\hat{H}_{\gamma_1} = & - \gamma_1 \sum_{i}\left(\hat{b}^{\dagger}_{1,i} \hat{a}_{2,i} + \mathrm{h.c.}\right)\nn
\hat{H}_{\gamma_3}=&-\gamma_3 \sum_{\langle i, j\rangle_3, }\left(a_{1 i }^{\dagger} b_{2 j }+\text { h.c. }\right)\nn
\hat{H}_{\gamma_4}=&\gamma_4 \sum_{\langle i, j\rangle_4}\left[a_{1 i }^{\dagger} a_{2 j }+b_{1 i }^{\dagger} b_{2 j }+\text { h.c. }\right]
\label{eq:H_1}
\end{align}
A creation (annihilation) operator of an electron at the atomic site $A_i$ is denoted as $\hat{a}^{\dagger}_{l,i}$ ($\hat{a}_{l,i}$) and similarly for the atomic site $B_i$ we use $\hat{b}^{\dagger}_{l,i}$ ($\hat{b}_{l,i}$). We sum over both layers $l=1,2$, each atomic site $i$ and also the nearest-neighbor hoppings $\langle i,j\rangle$. The first two terms denoted by $\hat{H}_0$ represent the onsite energy for each of the A and B atoms. Similarly, the hopping ($\gamma_0$) describes nearest-neighbor hopping within each graphene layer, ($\gamma_1$) couples the vertically aligned dimer sites ($B_1$) and ($A_2$), ($\gamma_3$) represents skew interlayer hopping between the non-dimer sites ($A_1$) and ($B_2$) and produces trigonal warping, while ($\gamma_4$) describes skew hopping between dimer and non-dimer sites across the two layers and introduces electron–hole asymmetry. In this model we do not include spin degree of freedom, assuming spinless electrons. The main reason is that we aim to study electron tunneling and single-particle Coulomb potential effects, and exclude weak spin-orbit coupling terms.

If we introduce an electrostatic bias between the layers the Hamiltonian term $\hat{H}_0$  in \eqref{eq:H_1} changes to 
\begin{align}
\label{eq:H_2}
\hat{H}_0 = &
\sum_{i}\frac{U(\vec{r})}{2}\left(\hat{a}^{\dagger}_{1,i}\hat{a}_{1,i}+\hat{b}^{\dagger}_{1,i}\hat{b}_{1,i}\right)
-\nonumber\\
&-\sum_{i}\frac{U(\vec{r})}{2}\left(\hat{a}^{\dagger}_{2,i}\hat{a}_{2,i}+\hat{b}^{\dagger}_{2,i}\hat{b}_{2,i}\right) .\end{align}
 
\begin{figure}[t]
    \centering

    \includegraphics[width=\columnwidth]{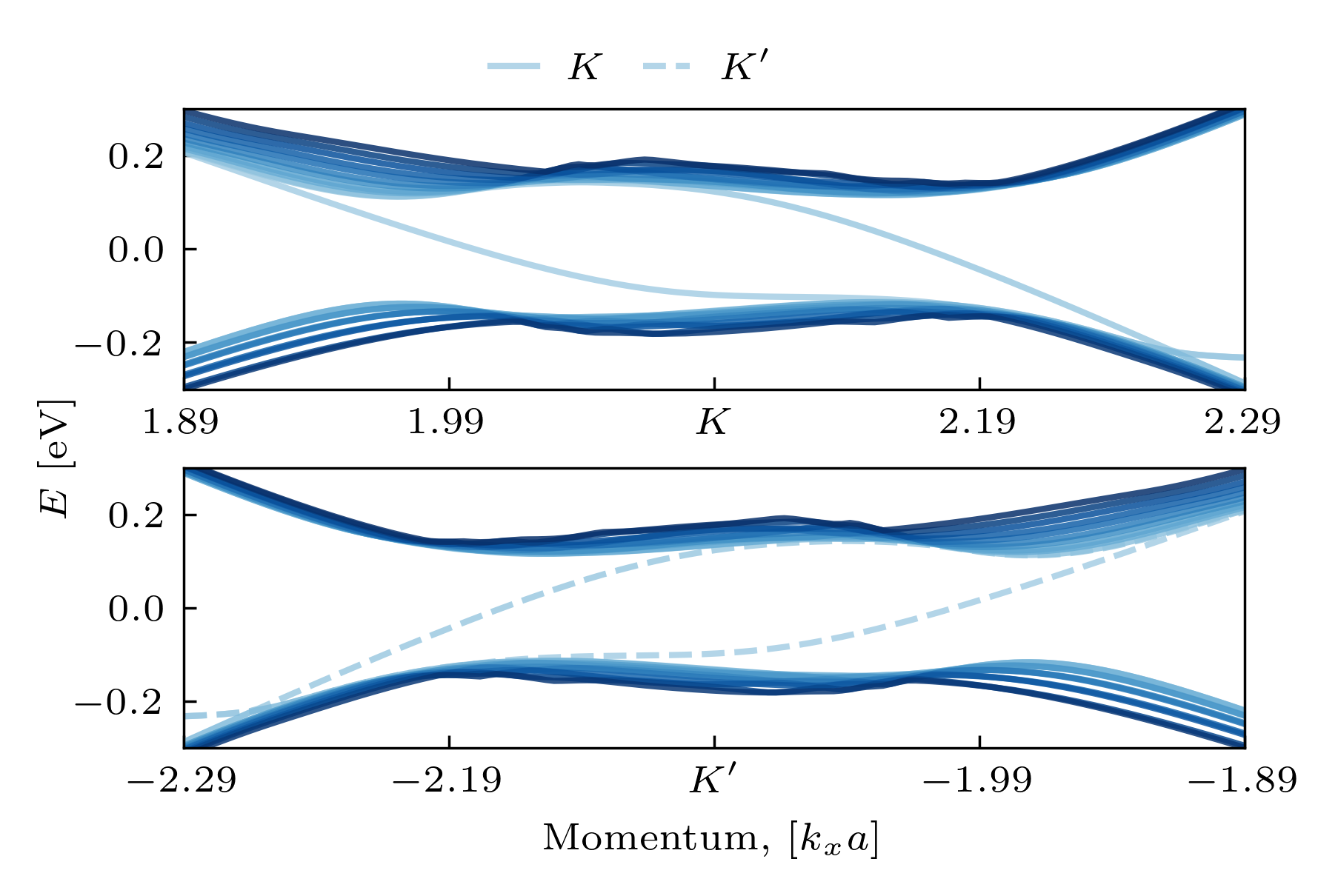}

    \caption{The band structure of BLG at each of the K-points when an interlayer bias is applied. From the Bulk-Boundary correspondence, one dimensional chiral modes appear in the band gap. The in-gap K valley modes are denoted with the solid blue lines, while the K' valley modes are the dashed blue lines.}
    \label{fig:K_valleys}
\end{figure}
In Eq. \ref{eq:H_2} the onsite energies change from $\epsilon^{A,B}_{l,i}$ values on atoms A and B to coordinate $\vec{r}$ and layer-dependent gate-induced bias. For the constant uniform $U(r)=U_0$ the band gap equals
\begin{align}
E_{\text {gap }}=\frac{|U_0| \gamma_1}{\sqrt{|U_0|^2+\gamma_1^2}}.
\end{align}
It saturates for a sufficiently high interlayer bias $U_0$ \cite{McCann2006PRB}. 

In the gapped Bernal bilayer graphene each valley acquires non-zero topological invariant, a so called valley Chern number \cite{Martin2008TopologicalConfinement,Zhang2013PNAS,Alden2013PNAS,Vaezi2013PRX}. The values of the valley Chern number are $+1$ in $K$-valley and $-1$ in $K'$-valley for a positive value of $U_0$. This value of valley Chern number ensures the existence of two edge modes per valley at the bias sign-change domain wall due to bulk-boundary correspondence. These modes are helical, moving in only one direction in each valley as the edge modes in quantum Hall effect. Thus, the corresponding transport phenomena is called quantum valley Hall effect, and was recently observed in Ref.\cite{Huang2024HighTemperatureQVH}.

\subsection{Helical valley Hall edge modes in domain wall}
Next, we discuss in detail the structure of these helical modes in the bias-defined domain wall. The domain wall of interest appears between the two regions having $+U_0$ and $-U_0$ value of inter-layer bias, respectively. Due to continuity and smooth nature of electrostatic potential, there is a region where effective inter-layer potential difference passes through zero. To study the localization of these modes, we model the spatial dependence of inter-layer bias in the quasi-one-dimensional channel aligned along x- or y-coordinate: 
\begin{align}
    U(x)=U_0 \tanh \frac{x}{\ell},\hspace{0.25cm}U(y)=U_0 \tanh \frac{y}{\ell}
    \label{eq:potential}
\end{align}
Here $\ell$ is the smoothening length of potential.  The leads with such structure are shown as horizontal elements of device in Fig.\ref{fig:top_down_device_schematic}.

In the Fig.\ref{fig:K_valleys} we show the numerically obtained spectrum of the channel near $K$ and $K'$ valleys. In the numerical simulation there is a subtlety: one needs to terminate the bilayer graphene edge in such a way that no non-helical edge modes appear in spectrum on the terminated ends. One way to overcome this was described in Ref.\cite{Luna2025SciPost} by introducing an effective hard wall boundary condition. Here we implement another way - a selection of particular atomic width modulo 3 of the ribbon, that does not have any edge states according to termination boundary conditions. Thus, both panels in Fig.\ref{fig:K_valleys} show the presence of only two helical modes per valley, being uni-directional. The used value of gap is $U/2=0.25 \,eV$.

\begin{figure}
    \centering
    \includegraphics[width=\columnwidth]{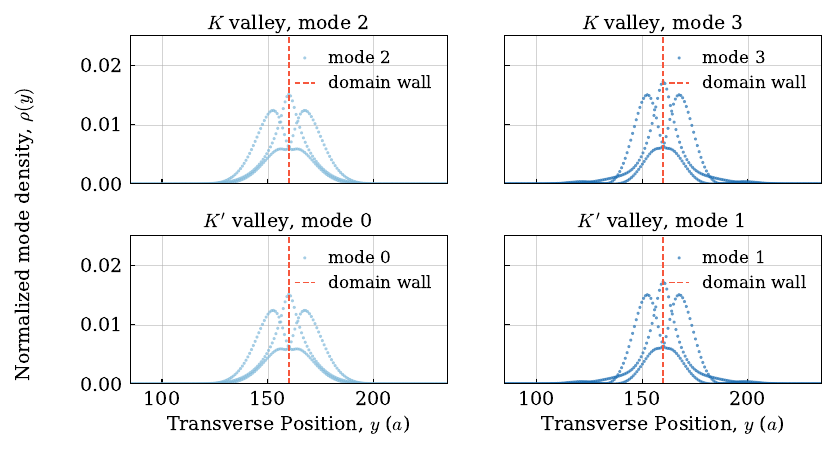}
    \caption{The transverse VHM profile along the electrostatic domain wall for each of the two helical modes present on either side. The electrostatic domain wall is denoted by the vertical red dashed line. What we see is that due to the unequal A and B lattice sites on either side of the domain wall, the wavefunction has two peaks on the inside and outside.}
    \label{fig:mode_profile}
\end{figure}

The spatial structure of the helical modes in quasi-1D channel in the cross-section slice is shown in Fig.\ref{fig:mode_profile}. In all four panels one finds that the localization peaks of the valley Hall modes are located on either side of the domain wall for the same valley. 
These peaks of the same mode are called the \textit{inner} and \textit{outer} parts in the scattering device.

\begin{figure}
    \centering
    \includegraphics[width=\columnwidth]{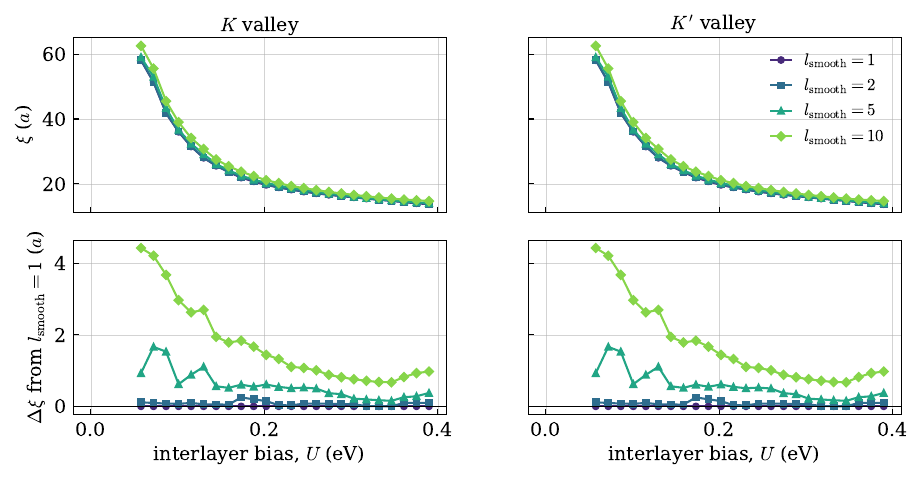}
    \caption{The evanescent decay lengths of the VHM for various interlayer biases and smoothening parameters. To make the domain wall sign shift more like a step-function, a smaller smoothening parameter ($I_{\text{smooth}}$) is chosen. Whereas a larger smoothening parameter indicates a longer bias sign transition. We see that for small interlayer biases (on the top row), we get larger decay lengths $\zeta (a)$. Additionally, for larger $I_{\text{smooth}}$, we get a longer decay length which is seen on the bottom row. $\Delta\zeta$ is the difference in decay length units ($a$) between the chosen smoothening parameter and $I_{smooth}=1$.}
    \label{fig:decay_length_parameters}
\end{figure}

The evanescent decay length of these modes are inversely proportional to the square root of the interlayer bias $\xi \sim \frac{2 \hbar v_F}{\sqrt{\gamma_1|U|}}$  \cite{Martin2008TopologicalConfinement,Zhang2013PNAS,Alden2013PNAS,Vaezi2013PRX}. This follows from the trivial argument of topological protection and dimensional analysis, as this is the only combination of dimension of length one could get in the system.
This means that for smaller inter-layer bias between the layers, the size of the evanescent decay length increases, which is shown in  Fig. \ref{fig:decay_length_parameters}. The value of the decay length can also increase slightly by changing the spatial variation of the electrostatic domain wall $\ell$. If there is a sharp, step-like domain wall, then the mode becomes highly localized along the edge. Furthermore, if the smoothening length $\ell$ is large, then the mode is not as localized and the decay length increases. For the purpose of reducing computational complexity, we select an interlayer bias of 0.25 eV. This value was chosen since it localizes the valley Hall modes well along the domain wall and also provides a relatively significant decay length size. Additionally, for a gap parameter of this size we only require a minimum system size of about 79 nm by 79 nm to have the valley Hall modes tunnel through the bulk. Since detailed fabrication of a device of this size can be challenging, it should also be noted that the system can be scaled up in size, as long as all proportionalities between values are held constant. In other words, as shown in the Fig.\ref{fig:decay_length_parameters}, in the system with smaller inter-layer biases the same scattering effects should appear at much larger spatial scales. This makes the predictions discussed in this paper being possible to reproduce in experimental devices of much larger atomic sizes, on the orders of hundreds - (for gap around $50$ meV) to thousands  nanometers (for gap around $1-5$ meV).

\subsection{External impurity potential}
Finally, we introduce the model for impurity potential. Impurity is used for two purposes: to verify the valley index conservation during the scattering at disorder and to study the effects of resonant tunneling of helical modes. In addition, while the valley index is protected by time-reversal symmetry, the flavor of two modes within one valley could be influenced by the impurity. Our aim is to describe this effect in  simulations under various impurity conditions. 

An impurity potential $V(x,y)$ is placed on one of the graphene layers and is shaped as a small potential well. The onsite potential at this impurity can be varied and also placed in different points in the scattering region. In the calculations below we put impurity in the center of scattering region between islands, as shown in Fig.\ref{fig:top_down_device_schematic}. This kind of impurity is most akin to screened Coulomb impurity, which is a charged particle (for example a nitrogen atom or vacancy \cite{Joucken2021,Zhou2022,Mao2016}) on the surface of the material. The typical charges for Coulomb impurities are integer multiples of the elementary atomic charge $Z=\pm k$ where $k \in \mathbb{Z}$. Due to screening effects of the electron gas in Bernal bilayer graphene \cite{Gamayun2011}, the $1/r$ drop off of the potential changes to a range potential with typical decay of $1/r^3$ and modified value of effective charge \cite{Skinner2014,Oriekhov2017PRB,Gorbar2024PRB}. If the impurity parameters are chosen correctly in the simulation, then it is able to host in-gap bound states. These in-gap states allow valley Hall modes to tunnel through the bulk of the scattering region resonantly. Other sizes and shapes of the impurity can be used, but for simplicity a single circular size of radius $R=5a$ was chosen. This size also reflects the possibility to have a small cluster of impurities with screened short-range potentials close together, as was used to experimentally observe atomic collapse \cite{Wang2013science_collapse}. In section \ref{sec:impurity}, more discussion will be centered around the calculated in-gap states of this impurity and the scattering effects it has on the valley Hall modes (VHM).




\section{Scattering through the gap}
\label{sec:scattering-between-islands}
\subsection{Clean system}
\begin{figure}
    \centering
    \includegraphics[width=\columnwidth]{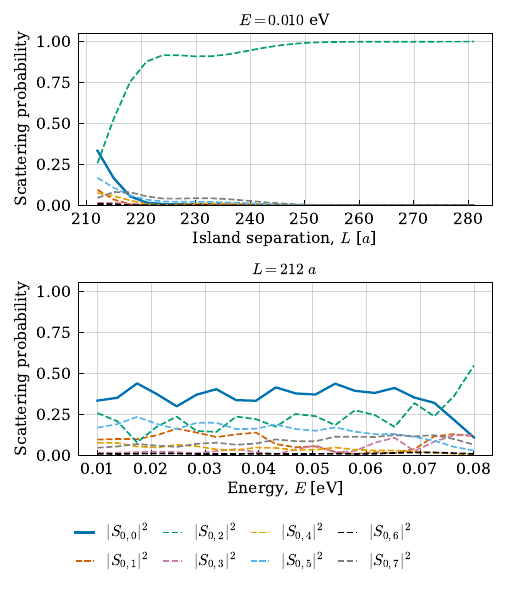}
    \caption{The length and energy sweep for the half circle system versus the scattering probability. For this plot, the mode originates from lead zero and can tunnel to seven other possible modes, before leaving the scattering region. At a length of 250$a$, there is an increased likelihood that the mode tunnels to other possible channels. Different energies at a fixed length, can enhance certain modes or suppress them.}
    \label{fig:1D_sweep}
\end{figure}
A simple model for the scattering of the helical edge modes is through a uniform bulk region, where there is a non-zero probability that the modes can tunnel across via evanescent decay to the available modes along another electrostatic domain wall. The scattering region is setup with two half-circles on either side and leads attached perpendicularity to the intersections of the domain wall with the vacuum (Fig.\ref{fig:top_down_device_schematic}). Injection of the modes are created in the lead and flows into the scattering region. Each lead can at most, host two incoming propagating modes, since the other two propagate out of the scattering region. As the mode traverses the domain wall, the maximal overlap of the electrons wavefunction occurs at the horizontal midpoint of the system. Higher rates of scattering through the bulk can occur at shorter length scales (more overlap) and at specific energies. Varying length scales and energies were simulated and their scattering matrices were recorded. Due to particle conservation, the scattering matrix elements of both the incoming and outgoing modes can be normalized to represent numerically the probability of tunneling to another available mode. As we expect, shortening the distance between the domain walls result in a higher tunneling probability. Similarly, various in-gap energies for fixed lengths exhibit higher tunneling probabilities. This can clearly be seen in the Fig. \ref{fig:1D_sweep}, where for the mode available to the electron originating in \textit{lead zero} (mode zero) in the scattering region with K-valley state, we plot its outgoing scattering probability as a function of distance and of energy. The indices of S-matrix elements $S_{ij}$ indicate the $i$-input index of mode and $j$-output. In the 4-lead geometry we have in total 8 modes incoming and 8 outgoing. If the mode scatters to the same lead from which it originated, it necessarily changes its valley index. The indices corresponding to each lead are indicated in schematic Fig.\ref{fig:blg-misaligned-patterns}(b).   The mode denoted $|S_{0,2}|$ represents the K-valley mode originating from \textit{lead zero} (bottom right corner), and terminates at \textit{lead one} (top right corner) in the available outgoing K-valley mode. This mode, at a sufficient distance, cannot traverse the bulk region and move to the available modes at the other domain wall. Rather this mode circles around the domain wall on the right side clockwise. As the distance between both semi-circles is decreased, there is an increase in the other available scattering probabilities which can be seen starting around the distance of 255$a$ between centers. This indicates that the probability to tunnel across the bulk region increases with the decreasing separation. Next, in the lower panel of Fig.\ref{fig:1D_sweep} we show the results of computation for the fixed inter-island distance and variable energy.  We see in lower panel of Fig.\ref{fig:1D_sweep} that for certain energies, there is an increase of the tunneling probability to certain modes available in the scattering region. 

As seen in upper panel in Fig. \ref{fig:1D_sweep}, in the energy sweep at a very close positions of islands the scattering probability is dominated by the $|S_{0,0}|^2$ matrix element. This element corresponds to an inter-valley scattering and return of mode back into the same lead. The semi-islands always have the radius of $100a$ in all simulated cases. If we pull the two half-circle islands further away, such that there is a larger distance between them, then we observe other matrix entries contributing to the scattering process. Additionally, a small particular oscillatory effect can be seen across the energy sweep, even though there is no clear periodic oscillation taken. At even larger lengths, such that the domain walls are sufficiently far apart, the energy sweep is dominated by $|S_{0,2}|^2$ and no oscillatory behavior is observed.  
This oscillatory behavior is related to the effects of finite-size quantization of modes confined around the island (see Ref.\cite{Benchtaber2021}). In addition, the suppression and re-appearance of tunneling in  scattering probabilities with growing distance was previously observed as an existence of parameters with fully prohibited scattering in Ref.\cite{Benchtaber2021scatt}. We show that this oscillatory behavior also depends on energy level of scattering mode. This could be seen in additional scattering matrix plots shown in Appendix \ref{sec:appendix-S-matrix-data}.
The scattering properties of the system becomes more non-trivial when an impurity with an in-gap bound state is present in the system.

\subsection{Impurity in the central region}
\label{sec:impurity}
\begin{figure}
    \centering
    \includegraphics[scale=1]{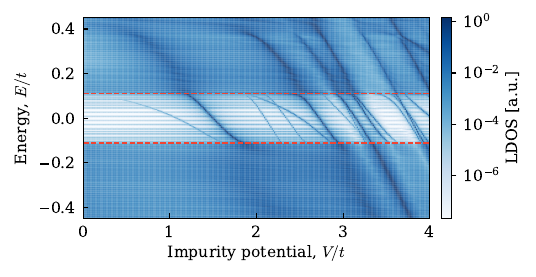}
    \caption{The local density of states (LDOS) of a central singular atomic site of a quantum dot impurity, as a function of the impurity potential and energy. The red dashed lines denote the opened band gap present due to a perpendicular interlayer bias. Bound states at the impurity are visualized by the dark blue, high LDOS branches that come from the bulk bands and appear within the band gap region.}
    \label{fig:LDOS}
\end{figure}
It is also possible to simulate an impurity in between both islands, which could host in-gap bound states the the VHM can tunnel to. Just like in the clean system, we vary the spatial overlap of the modes with the opposite domain wall and vary the energy. For each system of a fixed length and energy, the scattering matrix is recorded. The scattering matrices will allow us to model the tunneling probability to various available modes in the scattering region.

\begin{figure}
    \centering
    \includegraphics[scale=1]{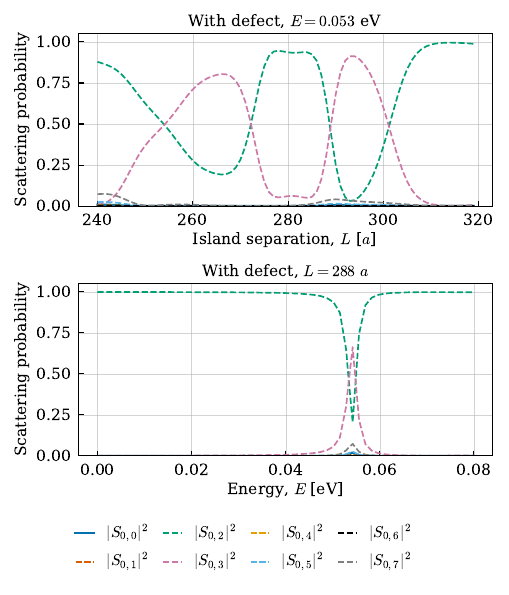}
    \caption{The length and energy sweep for the half-island system, but with a Coulomb impurity present. As the length of the system is varied, the valley flavor oscillates back and forth to preserve particle conservation. Other modes present in the system are significantly suppressed. Only at the resonant energy of the Coulomb defect, there is a sudden dip between valley flavors.}
    \label{fig:defect}
\end{figure}

When a simple impurity is present, in this case a circular Coulomb impurity with a radius of $R=5a$, the scattering properties of the system changes due to in-gap bound states. A sweep of the local density of states (LDOS) for the Coulomb impurity was taken for various onsite potentials and energies (see Fig. \ref{fig:LDOS}). The energy range is taken to measure both the bulk energy values and the in gap values. Dark blue branches, which represent larger LDOS values, move from the bulk spectrum to the band gap. In the gap, these branches host bound states that the VHM can tunnel to. Choosing one of the branch points, we can also plot the length and energy sweeps as before. What we see is that the Coulomb impurity causes a dramatic variability in the valley \textit{flavor} at it's resonant energy (see Fig. \ref{fig:defect}). This means that along the domain wall the valley mode shifts from the outer mode to the inner mode along the domain wall and vice versa. 

\begin{figure}
    \centering
    \includegraphics[scale=1]{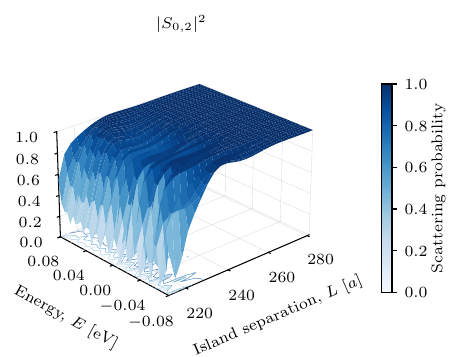}

    \caption{A three dimensional scattering probability plot as a function of the energy $E$ and island separation $L$. There is no Coulomb impurity present. Periodic dips as a function of energy can be seen just before the drop of the scattering probability. The weak periodic modulations preceding the sharp decrease are coherent interference between coupled Valley Hall domain wall modes. Changing \(E\) modifies both their accumulated phase and evanescent decay length, while decreasing \(L\) enhances their overlap, resulting in a energy and separation dependent redistribution of the outgoing channels.}
    \label{fig:3d_sweep}
\end{figure}

At other lengths and energies the scattering process of the VHM changes. We can visualize this change in a three dimensional plot of the lengths, energies versus the scattering probability. When the Coulomb impurity is introduced into the system, with its in gap bound state, we see that the VHM scattering probability changes dramatically. Fig. \ref{fig:3d_sweep} is showing only the case of an electron in mode zero tunneling to mode two, which is at the lead right above it. Compared to the clean system, where this tunneling process only changes at short enough distances between the domain wall, the tunneling out of this terminal mode is much more non-trivial with a Coulomb impurity present. It should be mentioned that most of the other modes for the impurity case did not see high scattering probabilities. The only significant scattering probabilities were between mode zero and mode two. What this means is that in the presence of an impurity with a in gap bound state, we do not see a significant increase in inter-valley scattering. 

\section{Effect of misalignment on edge mode and scattering between islands}
\label{sec:misalignment}
\begin{figure}
    \centering
    \includegraphics[scale=1.0]{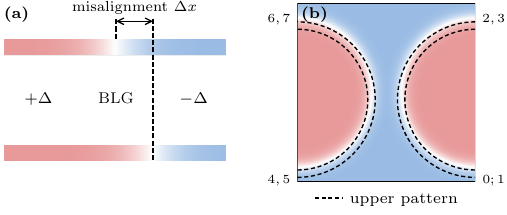}
    \caption{Schematic structure of misaligned leads and gate-defined islands. In addition, we show in panel (b) the number labeling of modes that are in-coming and out-going in S-matrix in the four-lead geometry.}
    \label{fig:blg-misaligned-patterns}
\end{figure}

The system that has been studied above assumes that there is a perfect alignment of the electrostatic domain wall on both graphene layers. In other words, gate-defined islands and the neutral barriers between gates are placed perfectly above each other on both sides of BLG. It is then easy to ask what are the effects of the VHM on a misaligned domain wall and does this affect the scattering of these modes? Small displacements from the perfectly aligned case were taken and can only reach a maximum displacement of the defined smooth length $\ell$ (see Eq.\eqref{eq:potential}). The structure of studied misaligned potential of gates in shown in Fig.\ref{fig:blg-misaligned-patterns}. We study both the spectrum of lead with top and bottom gates being displaced by $\Delta x$ and the two semi-island scattering geometry with semi-islands on top and bottom layer being displaced by a distance $\Delta y$. As the displacement of the domain wall with respect to both graphene layers increases, the interlayer bias becomes equal for both layers in a finite strip. Therefore, the allowed band gap, which is a result of an opposite interlayer bias, becomes smaller and smaller. This results in the band gap shortening and makes the VHM less localized along the domain edges \cite{Luna2025SciPost}.

\begin{figure*}
    \centering
    \includegraphics[scale=1.0]{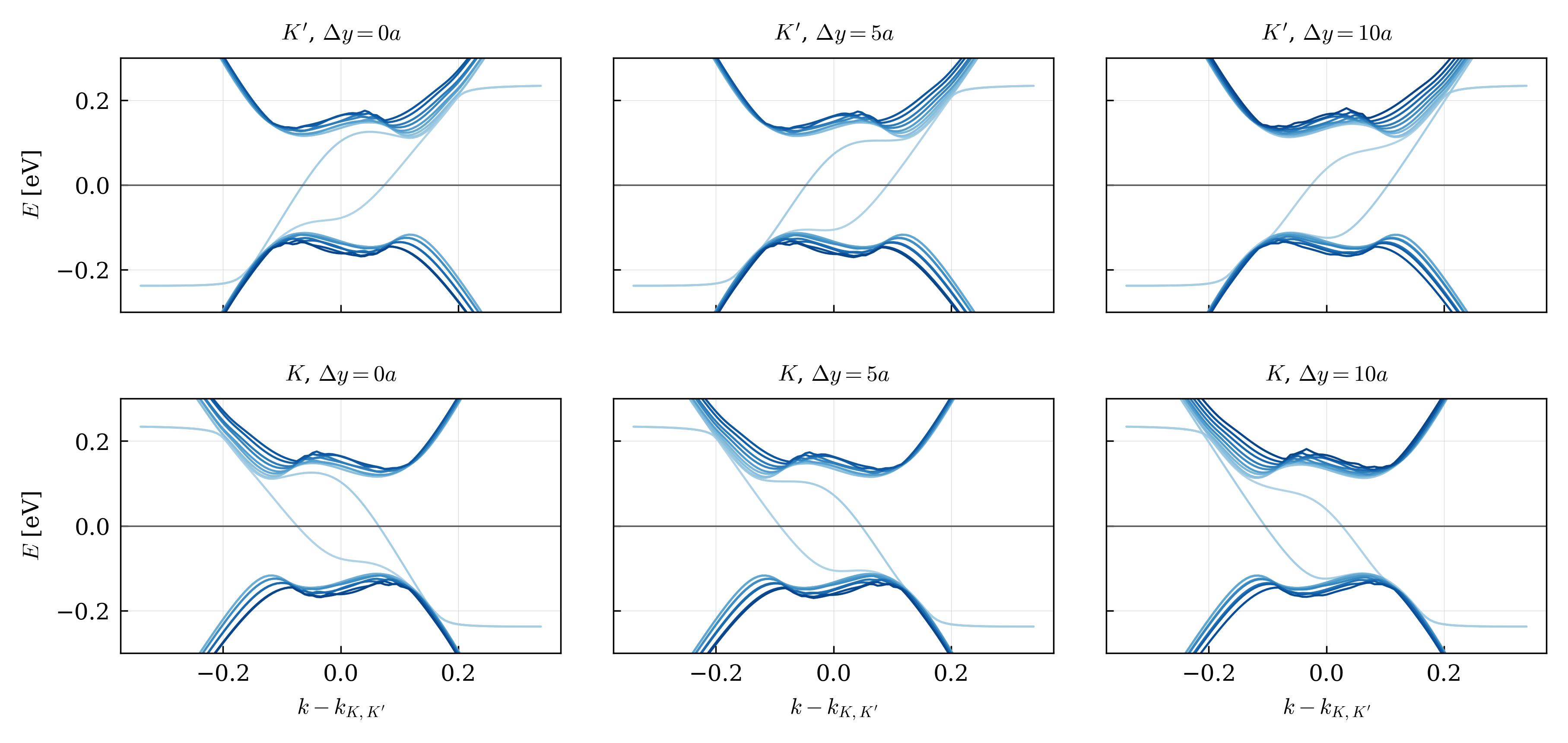}
    \caption{The band structure of the BLG system at the K and K' points for varying degrees of domain wall misalignment. There are three cases in this figure, the perfect alignment case (no mismatch; left), the half way misalignment case (mismatch is $5a$; middle), and finally the fully misaligned case (mismatch is $10a$; right). As the electrostatic domain wall gets more and more misaligned, the spectral curves of in-gap states shift in k-space, but maintain their general shape. The main effect, which was already pointed out in Ref.\cite{Luna2025SciPost}, is that the interval of equal Dirac speed of modes and their insulation from non-helical bands becomes displaced with respect to Fermi level.}
    \label{fig:band_structure_shift_DW}
\end{figure*}

\begin{figure*}
    \centering
    \includegraphics[scale=1.0]{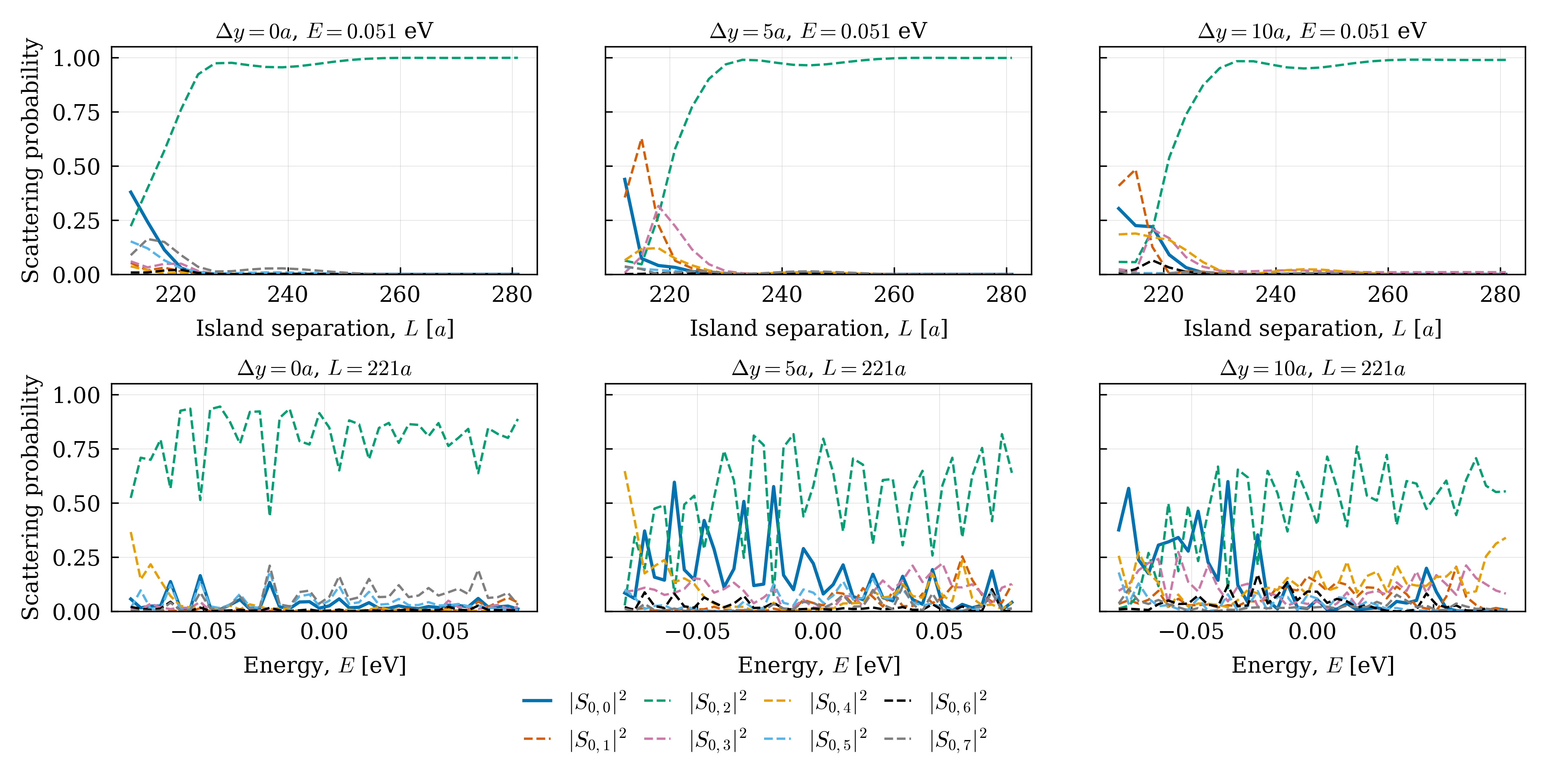}
    \caption{Dependence of scattering matrix elements absolute values on mismatch between upper and lower gates. Three scenarios are shown: left column - no mismatch, central column - mismatch $5a$ in each lead or $1/2$ of smoothening length $l$. And right column - mismatch 10a, or full smoothening length of potential $l$.}
    \label{fig:s_matrix_sweep_DW_misalignment}
\end{figure*}

Just as in Fig. \ref{fig:1D_sweep}, the scattering properties of the VHM for varying degrees of misalignment can be plotted for different lengths and energies. We see in Fig.\ref{fig:s_matrix_sweep_DW_misalignment} that compared to the perfectly aligned case, the scattering properties of the system do in fact change. Furthermore, the in gap states in Fig. \ref{fig:band_structure_shift_DW} shift for different misalignment values. The general trend of bulk scattering occurs (i.e. when the two domain walls become closer) even in the presence of misalignment. One noticeable difference is that other available modes, besides $|S_{0,2}|^2$, become more probable to scatter to when there is misalignment. The outer mode $|S_{0,2}|^2$ drops off quicker with misalignment as the inter-island distance decreases. Moreover, backscattering at the lead to bulk interface is quite large when the inter-island distance is short (modes $|S_{0,0}|^2$ and $|S_{0,1}|^2$ scatter back into the original lead). This does not seem to be the case when the inter-island separation is large, for example at $L=270a$. Sweeping through the energy also shows a difference in the presence of misalignment. In the perfectly aligned gate case, other modes besides $|S_{0,2}|^2$ are substantially suppressed, whereas in the aligned case these modes play a more active role. This is easily seen as the drop in the average scattering probability for mode $|S_{0,2}|^2$ between the allowed in gap energies. Additionally, the dominant channel $|S_{0,2}|^2$ drops significantly at lower energies, but not at higher energies. What all of this information tells us is that misalignment of the electrostatic domain wall is critical in reducing backscattering of the VHM into the bulk region. These effects are more noticeable at shorter inter-island distances than larger ones, but are also dependent at which in gap energy the VHM have.

\section{Splitting of energy levels in two islands geometry}
\label{sec:quantum-dot-simulator}

In this section we demonstrate a straightforward application of the obtained S-matrix properties. Namely, we suppose that one sample of bilayer graphene several gate-defined patters of round shape are placed in a close vicinity. Each gap-sign-alternating island represents a chiral quantum dot discussed in Refs.\cite{Xavier2010,Benchtaber2021}. We study the splitting of energy levels in such quantum dots due to tunneling. The similar system but only with gate-defined pattern on one side was already discussed in Ref.\cite{Zeng2024GateTunableTopologicalPhases}. In our case the modes around each patterned island are fully topologically protected.

The system that we simulate has the following geometry:
two islands of radius 100 are placed either far from each other or close nearby. According to S-matrices calculated in Sec. \ref{sec:scattering-between-islands}, specifically the amplitudes of its elements, we could expect the energy splitting of levels in latter geometry. We calculate the local density of states at the two places that are located on the edge of each island and sum up together. We again use kernel polynomial method (KPM) for this evaluation. We  compare the energy-resolved total LDOS at two point for separated islands, for close-by placed islands and the energy dependence of S-matrix at such separation.

The systems of such kind attract interest as potential host platform for various natural and artificial molecule simulators. This platform has an advantage of having very large radius of artificial atom and thus containing many possible mode states. In addition, the valley Hall modes are topologically protected compared to whispering gallery modes  \cite{brun2022graphene, seemann2024steering}, that are also angular harmonic modes in spatially-confined graphene devices. This could provide an alternative and larger size platform to tune artificial molecules compared to STM methods \cite{Sierda2023Science}. We leave the study of particular applications of such simulators for a future research, as it would require a careful treatment of electron-electron interactions within one helical channel.

\begin{figure*}
    \centering
    \includegraphics[scale=1.0]{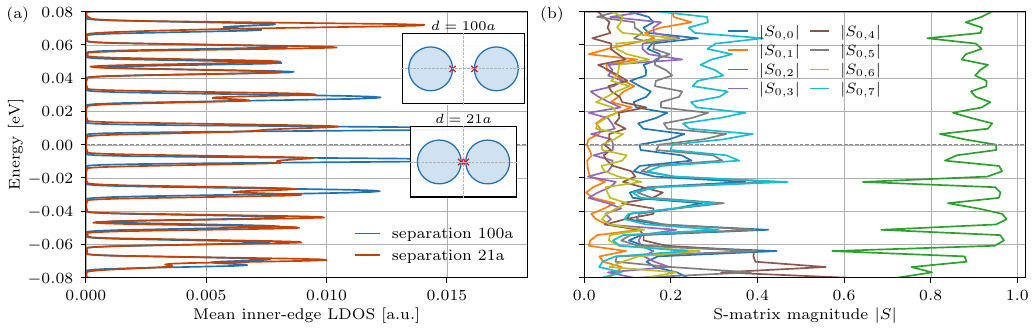}
    \caption{Comparison of KPM mean local density of states calculated at two edged of different islands, for systems at different separations. Inset in panel (a) demonstrates the two studied systems: islands at far distance and nearby on distance $21a$. The panel (b) shows S-matrix absolute values squared for the mode 0 injected from the right lead. The visible level splitting in the nearby islands geometry corresponds to nontrivial scattering peaks to another island in panel (b). At the same time, for suppressed scattering around $E=-0.06 eV$ one finds no level splitting despite high LDOS of states at each island. Radius of islands is $100a$.}
    \label{fig:kpm-molecule-test}
\end{figure*}

\section{Conclusion and Outlook}
 In the present paper we investigated the scattering and tunnel coupling properties of valley Hall domain wall modes in a double-gate-patterned Bernal bilayer graphene.
  The calculations were performed using the tight-binding lattice simulations with actual parameters of BLG lattice \cite{McCannKoshino2013}. We systematically investigated the following properties that affect in-gap scattering: localization of chiral modes in domain walls, dependence on inter-layer bias, the distance between domain walls, the Fermi level and the effect of impurity with local potential.
   The results of this work connect the microscopic properties of domain walls Bernal bilayer graphene with the effective scattering process needed for a larger network of coupled valley Hall channels. 

We first characterized the localization of the domain wall modes and found that their evanescent decay length is controlled primarily by the interlayer bias and, to a lesser extent, by the smoothness of the electrostatic domain wall. The larger the interlayer bias, the more strongly localized are the valley Hall modes. Smoothening the spatial extent of the electrostatic domain wall increased the evanescent decay length of these modes. As a result, we can modulate the tunnel coupling between neighboring channels either geometrically or electrostatically. The tunneling probability increases as the geometrical distance between domain walls decreases due to the spatial overlap of the wavefunctions. At sufficiently large distances between domain walls, the Valley Hall modes follow the path of their original trajectory along domain wall. All these results well agree with the previous findings in literature on exponential decay of valley Hall modes \cite{Martin2008TopologicalConfinement,Alden2013PNAS,Zhang2013PNAS,Xavier2010,Benchtaber2021scatt,Luna2025SciPost}. At the same time, our main contribution is the detailed study of S-matrix properties with varying external parameters. This provides a base for the simulation of setup with impurity and a few islands. In addition, it clearly confirms the observation about fully prohibited scattering at certain doping levels and distances between channels made in Ref.\cite{Benchtaber2021scatt}.

Introducing an impurity which can support an in-gap bound state changes the relatively smooth tunneling behavior. If the mode is at the resonant energy of the impurity bound state, the tunneling is highly dependent on the geometrical spacing between domain walls. The mode flavor within one valley, meaning whether the valley Hall mode is on the inner side or outside of the domain wall, adds extra dependence to this tunneling. An impurity with an in-gap bound state preserves the valley index and therefore significantly suppresses inter-valley scattering. These impurities can act as resonant couplers between spatially separated valley Hall modes without necessarily destroying the valley index and make higher tunneling rate between distant islands. This represents an important caveat that could influence the future experiments with valley Hall modes as carries of quantum information.

Additionally, the relative displacement between the electrostatic domain walls on each graphene layer was studied. The in-gap band structure remains qualitatively similar under moderate misalignment, which agrees well with findinds of Ref.\cite{Luna2025SciPost}, but the corresponding scattering properties are more sensitive to this variation. Misalignment increased the amount of backscattering and inter-valley scattering, especially for the shorter spacing between domain walls. As a result, in fabrication process the alignment of these domain walls should be considered as an important parameter for coherent transport, especially for the network models. At the same time, the relative stability of modes even at high misalignment values makes multi-island devices realistic to be fabricated with modern lithography techniques.

Finally, we studied the local density of states in the domain walls of two-island device. We found that the level splitting is generally present for close-by placed islands as a consequence of standard quantum mechanical lift of degeneracy by tunneling. At the same time, we also underline the role of suppressed scattering: at certain parameters - energy, distance  and interlayer bias, the energy levels of in-gap states do not experience splitting. This appears due to suppressed scattering found in the S-matrix. Leveraging such scenario, many islands could be placed together and host almost perfectly protected states insulated from each other. 

The systems studied in this paper assume a coherent, spinless, single-particle tight binding description of electron transport. Additionally, the electrostatic potentials and impurity potentials are assumed to be uniform and unchanging. A future work should be aimed to combine the described single-particle effects with electron-electron interactions, charge loading to domain walls and the question of formation of exotic states in domains walls, such as superconductivity \cite{Barrier2024} found in deformation domain walls, or Wigner crystals. In addition, we elaborate on the possible sizes of islands that should be available in experimental devices. The BLG system that has been discussed extensively in this paper can be scaled up to larger sizes by lowering interlayer bias, so long as all the proportionalities are kept constant. In particular, all observed effects should be observable on the order of system sizes of 5 to 10 times larger, by lowering the gap to $1$ to $5$ meV scale. That being said, the base size of the simulated square system is about 80 by 80 nm. This results in an area of $6,400$ nm$^2$. The radius of the superlattice islands is approximately 25 nm, which results in an area of $\sim 1,960$ nm$^2$. Since the valley Hall modes have been shown experimentally to survive at length scales on the order of a few micrometers \cite{Huang2024HighTemperatureQVH}, this means that the system can be scaled up by an order of magnitude \cite{Ju2015TopologicalValleyTransport}. Current experimental capabilities should allow for the fabrication of such a device in the lab.\\

\begin{figure*}[t]
    \centering
    \begin{minipage}{0.32\textwidth}
        \centering
        \includegraphics[width=\linewidth]{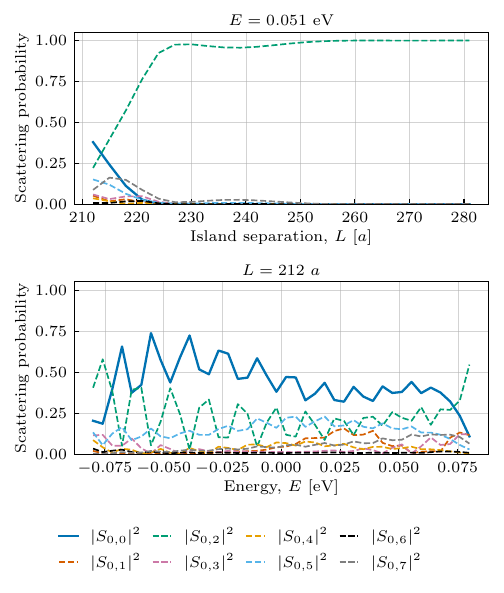}
    \end{minipage}
    \hfill
    \begin{minipage}{0.32\textwidth}
        \centering
        \includegraphics[width=\linewidth]{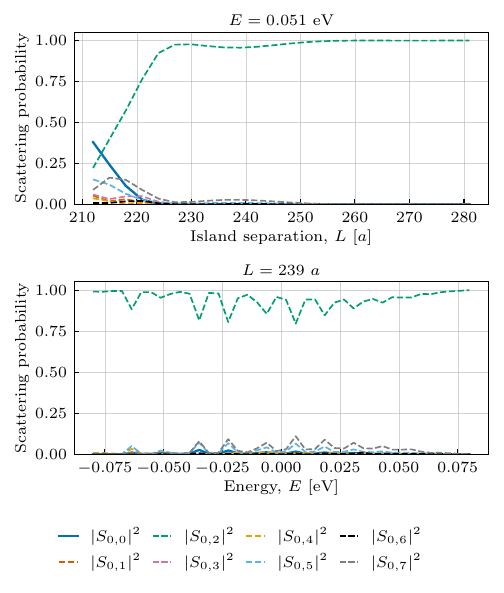}
    \end{minipage}
    \hfill
    \begin{minipage}{0.32\textwidth}
        \centering
        \includegraphics[width=\linewidth]{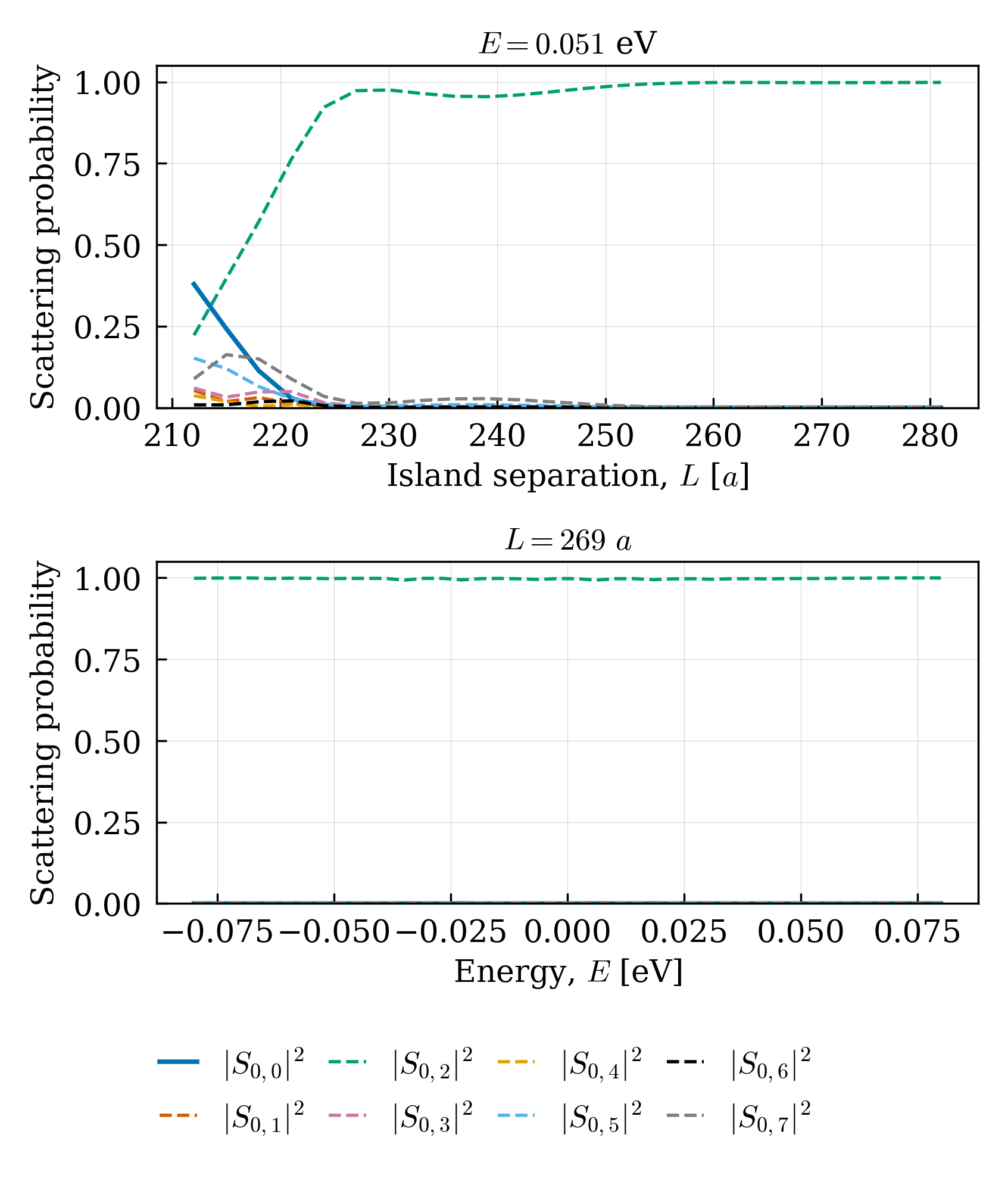}
    \end{minipage}

    \caption{Multiple energy sweeps for different island separation distances, ranging from close ($212a$), to medium ($239a$), to far ($269a$). The sweeps over the island separations are the same (top row), since we are viewing them from the same energy $E$. At shorter inter-island distances, the larger the oscillatory tunneling effects of the scattering probability. At far island separation distances (bottom right), there is no visible oscillation effect that takes place.}
    \label{fig:Extra_Sweeps_positive_E}
\end{figure*}

\begin{acknowledgements}
We acknowledge fruitful discussions with Johanna Zijderveld, Gonçalo Meneses, Anton Akhmerov, Eliska Greplova, Kostas Vilkelis, Isidora Araya Day. A.L.R.M.  acknowledges the funding from the European Research Council (Grant Agreement No. 856526).
D. O. O. acknowledges the support by Kavli Foundation. 
The authors acknowledge the use of computational resources
 of the DelftBlue supercomputer, provided by Delft High Performance Computing Centre \cite{DHPC2024}.

{\it AI usage disclosure}
We have used OpenAI Codex with
the GPT-5.5 Codex and GPT-5.6 Sol models to set up several parts of the numerical simulations and plotting. All AI-generated code was verified and independently tested by human.
\end{acknowledgements}

\begin{center}
    {\bf Data availability statement}
\end{center}
The code and generated data for all simulations performed
in the paper can be found in Ref. \cite{Supplement-code}.

\appendix
\section{Supplementary Results on other Parameter Regimes for S-Matrix}
\label{sec:appendix-S-matrix-data}




Here in the appendix we provide additional sweeps for a range of island separations ($L$) and energies ($E$). What is important to note here is that as the inter-island separation increases, the oscillatory effects in the energy sweeps reduces. This can be seen in the energy sweeps of Fig.\ref{fig:Extra_Sweeps_positive_E}. This phenomena is due to the finite-sized quantization of the valley Hall modes around the island. As the spatial overlap increases, there are certain energies where the phase coherence between both available modes matches exactly. This is where we see oscillating peaks (greater tunneling), followed immediately by dips in the energy sweep spectrum.



\newpage
\bibliography{graphene_qrw_bib}


\end{document}